\documentstyle[twoside,fleqn,espcrc2]{article}

\newcommand{\AmS}{{\protect\the\textfont2
  A\kern-.1667em\lower.5ex\hbox{M}\kern-.125emS}}

\title{Topology in SU(2) Yang-Mills theory\thanks{Partially supported
by EC Contract CHEX-CT92-0051 and by MURST.}}

\author{B. All\'es\address{Dipartimento di Fisica, Sezione Teorica,
   Universit\`a degli Studi di Milano and INFN, \\
   Via Celoria 16, 20133 Milano, Italy.}\thanks{Speaker at the conference.},
   M. D'Elia\address{Dipartimento di Fisica, Universit\`a di Pisa and INFN, \\
   Piazza Torricelli 2, 56126 Pisa, Italy.},
   A. Di Giacomo$^{\rm b}$ and R. Kirchner$^{\rm b}$}
       
\begin{document}

\begin{abstract}
New results on the topology of the $SU(2)$ Yang-Mills theory
are presented. At zero temperature we obtain the value of 
the topological susceptibility by using the recently introduced
smeared operators as well as a properly renormalized geometric 
definition. Both determinations are in agreement. At non-zero 
temperature we study the behaviour of the topological susceptibility 
across the confinement transition pointing out some qualitative 
differences with respect to the analogous result for the $SU(3)$ 
gauge theory.
\end{abstract}

\maketitle

\section{INTRODUCTION}

A relevant quantity to understand the breaking of the 
$U_A(1)$ symmetry in QCD is the topological susceptibility
$\chi$ in pure Yang-Mills 
theory~\cite{witten,veneziano}
\begin{equation}
\chi \equiv \int d^4 x \langle 0 | 
T(Q(x) Q(0) ) | 0 \rangle_{\rm quenched}, 
\label{qchi1}
\end{equation}
where
\begin{equation}
Q(x) = {g^2 \over 64 \pi^2} \epsilon^{\mu\nu\rho\sigma}
\label{qchi}
F^a_{\mu\nu}(x) F^a_{\rho\sigma}(x)
\label{qchi2}
\end{equation}
is the topological charge density.

The prediction is~\cite{witten,veneziano}
\begin{equation}
\chi = {f_\pi^2 \over 2 N_f} \left(
m_\eta^2 + m_{\eta'}^2 - 2 m_K^2 \right) \approx (180 \;\hbox{MeV})^4
\label{180}
\end{equation}
where $N_f$ is the relevant number of flavours. Eq. (\ref{180}) implies 
a well defined prescription  \cite{witten,veneziano} to deal with 
the $x \rightarrow 0$  singularity in eq. (\ref{qchi1}).

In \cite{noi} $\chi$ was evaluated at zero and
finite temperature for $SU(3)$ Yang-Mills theory. The value obtained
at zero temperature was in agreement with the 
prediction of Eq.~(\ref{180}). 
The value of $\chi$ at finite temperature displayed a sharp drop 
beyond the deconfinement transition. Here we give a short review
of a similar calculation for the $SU(2)$ gauge group \cite{noi2}. In addition 
we discuss the comparison with the geometric method, and we show 
that, after a proper renormalization of the latter, the two procedures give
consistent results \cite{noi3}.

\section{RENORMALIZATIONS}

Let $Q_L(x)$ be any definition of the topological charge density 
on the lattice. 
The lattice topological susceptibility $\chi_L$ is defined as
\begin{equation}
\chi_L \equiv \langle \sum_x  Q_L(x)  Q_L(0) \rangle .
\label{chiL}
\end{equation}
The lattice regulated $Q_L(x)$  is related to the 
continuum $\overline{MS}$ $Q(x)$ by a finite renormalization \cite{haris}
\begin{equation}
Q_L(x) = Z(\beta) Q(x) a^4 + O(a^6)
\label{Z}
\end{equation}
where $\beta\equiv 2 N_c/g^2$ in the usual notation.

Since $\chi_L$ does not obey in general the prescription~\cite{witten,veneziano} leading to eq. (\ref{180}), besides the 
multiplicative renormalization of eq. (\ref{Z}) there is also an additive 
renormalization
\begin{equation}
\chi_L = Z(\beta)^2 \chi a^4 + M(\beta) + O(a^6)
\label{M}
\end{equation}
where $M(\beta)$ contains mixings with operators of dimension
$\leq 4$.

In order to extract the physical signal
$\chi$ from eq. (\ref{M}), we need a determination  of the renormalization constants $M$ and
$Z$. We will determine them using the non--perturbative method of ref. \cite{vicari,gunduc,noi}. 

\section{THE MONTE CARLO SIMULATION}

We have used Wilson action and the usual heat-bath updating algorithm.
The scale $a(\beta)$ was fixed by using the results of ref. \cite{a,b}.

\subsection{Zero Temperature}

The simulation was done on a $16^4$ lattice.

We have used various definitions for $Q_L$. The $i$-smeared field theoretical 
$Q_L^{(i)}(x)$ is defined as
\begin{eqnarray} 
&&Q_L^{(i)}(x) = {{-1} \over {2^9 \pi^2}} 
\sum_{\mu\nu\rho\sigma = \pm 1}^{\pm 4} 
{\tilde{\epsilon}}_{\mu\nu\rho\sigma} \times \nonumber \\
&& \; \; \; \; \; \; \; \; \; \; \; \; \; \; \hbox{Tr} \left( 
\Pi^{(i)}_{\mu\nu}(x) \Pi^{(i)}_{\rho\sigma}(x) \right).
\label{qsmear}
\end{eqnarray}
$\Pi^{(i)}_{\mu\nu}$ is the plaquette in the $\mu - \nu$ plane
constructed with $i$-times smeared links $U_\mu^{(i)}(x)$~\cite{smear}. 
We call $M^{(i)}$ and $Z^{(i)}$ the
additive and multiplicative renormalization constants for the
$i$-smeared operators. Of course $\chi$ must be independent of the choice of 
the operator $Q_L$.

\begin{figure}[htb]
\vspace{4.5cm}
\includegraphics{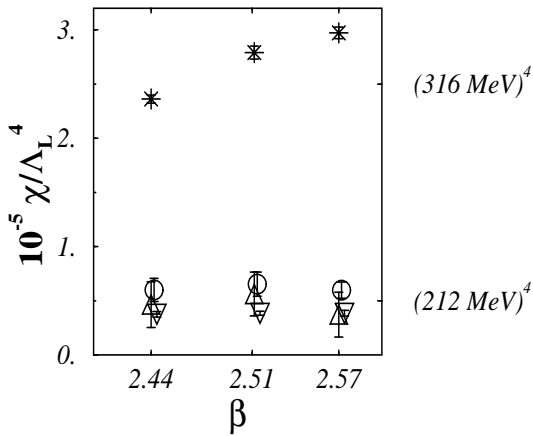}
\null\vskip 0.3cm
\caption{Topological susceptibility at zero temperature.} 
\end{figure}

This is visible in Figure 1, for $\chi$ at zero temperature.
Up and down-triangles indicate the value
of $\chi$ as obtained from the 0-smear and 2-smear data respectively.
There is good scaling and $(\chi)^{1/4}=(198\pm 2\pm 6)$ MeV,
the first error is statistical and the second comes from the
error in $\Lambda_L$~\cite{a,b}.

In Fig. 1 we also report the geometric susceptibility
$\chi_L^{\rm g}$ \cite{geom1,geom2}. The usual determination is done by 
identifying $\chi_L^{\rm g} = \chi a^4$, and claiming that the geometrical
susceptibility has no additive renormalization. $\chi_L^{\rm g}$ is shown in 
Figure 1 by the stars: it is one order of magnitude bigger than the field 
theoretical determinations and no scaling is observed.
By using the same method as for the field theoretical 
determination, we have measured the multiplicative
renormalization $Z^{\rm g}$, finding $Z^{\rm g} = 1$ within errors.
At the same time we have determined and subtracted the additive 
renormalization $M^{\rm g}$. This brings down the resulting $\chi$ by a factor 
of 10. The results are shown by the circles in Figure 1 and are in agreement 
with the field theoretical results. By renormalizing we have 
eliminated the so--called dislocations.

\begin{figure}[htb]
\vspace{4.5cm}
\includegraphics{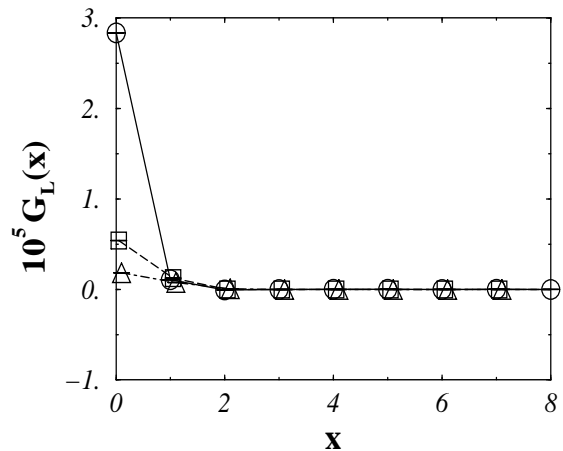}
\null\vskip 0.3cm
\caption{Correlation function for the $i$-smeared charges at $\beta=2.57$.
The lines are to guide the eye.
$i$=0,1,2 correspond to circles, squares and triangles respectively.}
\end{figure}

\begin{figure}[htb]
\vspace{4.5cm}
\includegraphics{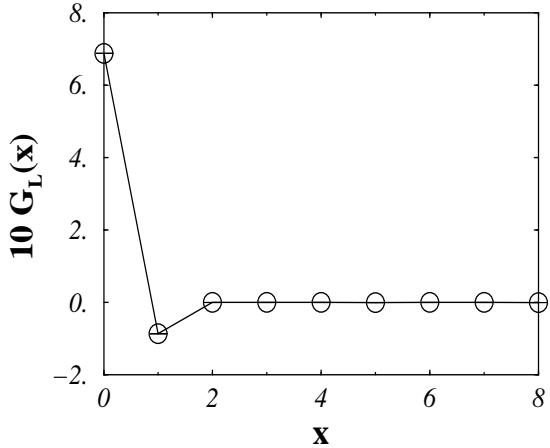}
\null\vskip 0.3cm
\caption{Correlation function for the geometrical charge at $\beta=2.57$.
The line is to guide the eye.} 
\end{figure}

To support the necessity of the subtraction of $M^{\rm g}$, 
we have also computed the correlation function
\begin{equation}
G_L(x) \equiv \langle Q_L(x) Q_L(0) \rangle
\label{G}
\end{equation}
By reflection positivity we expect that $G_L(x) \leq 0$ at
$x \neq 0$. 
Since $\langle Q^2_L \rangle > 0$, the susceptibility is mainly determined 
by the singularity at $x = 0$. 
This correlator, for $x$ lying along a coordinate axis,
is shown in Figures 2 for the smeared charges and in Figure 3
for the geometrical charge.
The peak at $x = 0$ for the
geometric charge is $4-5$ orders of magnitude larger than for the 
$i$-smeared charges, indicating that $M^{\rm g}$ is much bigger than 
$M^{i}, i = 0,1,2$, and is $\sim 80 \%$ of the observed $\chi_L$.



\subsection{Finite Temperature}

The simulation was done on a $32^3\times 8$ lattice. At this
size the deconfining transition is located 
at $\beta_c=2.5115(40)$~\cite{b} which means that 
$T_c=1/(N_t a(\beta_c))$ with $N_t$ the temporal size of the lattice.

\begin{figure}[htb]
\vspace{4.5cm}
\includegraphics{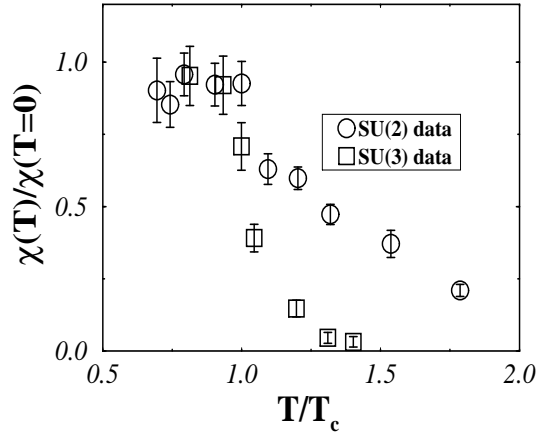}
\null\vskip 0.3cm
\caption{Ratio $\chi(T)/\chi(T=0)$ for $SU(3)$ and $SU(2)$.}
\end{figure}

The data show again a drop at the transition. However this is less
sharp than for the $SU(3)$ case \cite{noi,noi2}. In Figure 4 we show
the behaviours for $SU(2)$ and $SU(3)$ of the ratio 
$\chi(T)/\chi(T=0)$, where $\chi(T)$ indicates the physical susceptibility
at temperature $T$. The slope for the $SU(3)$ data is steeper.
In both cases the data at $T < T_c$ show a constant value 
consistent with the value at $T=0$.

\vskip 5mm

We thank Prof. Gerrit Schierholz for providing us with the fortran code
for the geometrical charge.


\end{document}